\documentclass[twocolumn,prb,showpacs,citeautoscript,floatfix]{revtex4}
\usepackage{graphicx}

\begin{document}

\title{Pressure-induced phase transitions in the multiferroic perovskite BiFeO$_3$ studied by 
far-infrared micro-spectroscopy}

\author{A. Pashkin}
\author{K. Rabia}
\author{S. Frank}
\author{C. A. Kuntscher}
\email[E-mail:~]{christine.kuntscher@physik.uni-augsburg.de}
\affiliation{Experimentalphysik 2, Universit\"at Augsburg, D-86135
Augsburg, Germany}
\author{R. Haumont}
\author{R. Saint-Martin}
\affiliation{Laboratoire de Physico-Chimie de l'Etat Solide
(CNRS),Universit\'e Paris XI, 91405 Orsay, France}
\author{J. Kreisel}
\affiliation{Laboratoire Mat\'eriaux et G\'enie Physique (CNRS),
Grenoble Institute of Technology, 38016 Grenoble, France}

\date{\today}

\begin{abstract}
We present the results of pressure-dependent far-infrared reflectivity
measurements on the multiferroic perovskite BiFeO$_3$ at room
temperature. The observed behavior of the infrared-active phonon modes
as a funtion of pressure clearly reveals two structural phase transitions
around 3.0 and 7.5 GPa, supporting the results of recent Raman and
x-ray diffraction studies under pressure. Based on the
pressure-dependent frequency shifts of the infrared-active phonon modes 
we discuss the possible character of the phase transitions.
\end{abstract}

\pacs{77.80.-e, 75.50.Ee, 78.30.-j, 62.50.-p}

\maketitle

\section{Introduction}

The perovskite BiFeO$_3$ is a robust magnetoelectric
multiferroic,\cite{Smolenskii82,Fiebig05} with the coexistence of
ferroelectric and antiferromagnetic order up to unusually high
temperatures. BiFeO$_3$ bulk samples exhibit an antiferromagnetic
N\'eel temperature of $\sim$370$^\circ$C and a ferroelectric Curie
temperature of $\sim$830$^\circ$C.\cite{Kiselev63,Smolenskii61}
For possible applications the growth of high-quality thin films
and the study of their physical properties are of major interest.
It is known that the properties of multiferroic thin films can be
significantly altered by the lattice mismatch between the
material and the substrate.\cite{Wang03,Li04,Yun04,Prellier05}
The enhanced ferroelectric polarization in BiFeO$_3$ films
\cite{Wang03,Li04} compared to bulk material were initially proposed 
to be driven by epitaxial strain; this matter is, however,
controversially debated.\cite{Eerenstein05,Wang05,Ederer05,Bea06}
Recently a high spontaneous polarization value - close to
theoretical predictions \cite{Neaton05,Ravindran06} - has been
also reported for high-quality BiFeO$_3$
ceramics\cite{Shvartsman07} and single
crystals,\cite{Lebeugle07,Lebeugle07a} demonstrating that a large
spontaneous polarization is an intrinsic property of BiFeO$_3$
bulk samples and probably not induced by a strain.

In general, BiFeO$_3$ is a complex system with magnetic, ferroelectric and
ferroelastic order parameters which are mutually coupled. Thus,
various instabilities can be driven by external thermodynamical
variables, like temperature, pressure, electric or magnetic field,
resulting in a particularly rich phase diagram. 
Experimental\cite{Haumont06,Kreisel07a,Gavriliuk05,Gavriliuk07}
and theoretical\cite{Ravindran06} investigations have investigated the
influence of external pressure on BiFeO$_3$. In fact, the importance of
high-pressure studies was demonstrated for a number of
ferroelectric
materials.\cite{Venkateswaran98,Wu05,Kreisel02,Chaabane03,Kornev05}
\textit{Ab initio} calculations of the total energy for different
structural arrangements of BiFeO$_3$ suggest that at pressures
above 13~GPa the $Pnma$ phase possesses lower energy than the
$R3c$ phase.\cite{Ravindran06} Thus, a pressure-induced structural
phase transition from the polar rhombohedral $R3c$ structure to
the nonpolar orthorhombic $Pnma$ structure was predicted. 
Pressure-dependent Raman and x-ray diffraction studies carried out
on BiFeO$_3$ single crystals\cite{Haumont06,Kreisel07a} 
indeed revealed two pressure-induced structural phase transitions at
around 3 and 10~GPa. The first phase transition at
$P_{c1}\approx$3~GPa was assigned to a distortion of the BiFeO$_3$
rhombohedral unit cell; however, the exact character of the
structural changes could not been determined yet. The second phase
transition at $P_{c2}\approx$10~GPa is most probably related to a suppression 
of the cation displacements (with a concomitant suppression of the
ferroelectricity), and it was proposed that the crystal structure
changes from rhombohedral $R3c$ to orthorhombic $Pnma$, in good
agreement with recent ab-initio calculations\cite{Ravindran06}
(although the experimental work suggests an intermediate bridging
phase).

The low pressure of the first phase transition ($\approx
3$~GPa) in bulk BiFeO$_3$ indicates a high sensitivity of the
system regarding stress and may originate from a complex
interplay between the magnetic, ferroelectric and ferroelastic
order parameters. Furthermore, experimental studies at very high
pressures\cite{Gavriliuk05,Gavriliuk07} (up to 70~GPa) have
revealed a transition from an antiferromagnetic to a nonmagnetic
state at 47~GPa. In the same pressure range BiFeO$_3$
undergoes an insulator-to-metal transition evidenced by optical
and transport measurements.\cite{Gavriliuk07}

Until now the reported far-infrared reflectivity measurements on BiFeO$_3$
addressed only the temperature dependence of the phonon
response. Kamba et al.\cite{Kamba07a} reported far-infrared reflectivity
spectra of BiFeO$_3$ ceramics in a broad temperature range (20 -
950~K) and discovered a softening of some phonon modes on
approaching the ferroelectric transition temperature. Recently,
far-infrared spectra of single
crystals\cite{Lebeugle07,Lebeugle07a} between 5~K and 300~K have
been presented by Lobo et al.\cite{Lobo-arxiv} with careful
assignment of the phonon modes and analysis of their contribution
to the static dielectric constant.

In this paper we report the effect of pressure on the far-infrared
response of BiFeO$_3$ single crystals at room temperature.
Reflectivity spectra of single-crystal samples were measured using
far-infrared micro-spectroscopy in combination with a diamond
anvil high pressure cell. The primary motivation of our
investigations was to confirm the recently found structural phase
transitions under high pressure and to obtain additional
information about their character.

In general, high-pressure infrared studies of phonon modes are
rare in the literature, mainly because of the experimental
difficulties when compared with Raman scattering. To the best of
our knowledge, the present work is the first systematic study of the
phonon behavior in ferroelectrics under high pressure by means of
infrared reflection spectroscopy.

\section{Experiment}
\label{sectionexperiment}

The investigated BiFeO$_3$ single crystals were grown using a $\rm
Fe_2O_3 / Bi_2O_3$ (1:4 M ratio) flux in a platinum crucible. The
flux was held at 920$^\circ$C and slowly cooled, similarly to the
previously reported procedure.\cite{Kubel90} Light yellow
translucent crystals in a shape of thin platelets have been
isolated by dissolving the flux in dilute nitric acid.
Back-reflection Laue photographs indicate a (001)$_{pc}$
crystallographic orientation of the platelets (where the index
$pc$ denotes a pseudo-cubic setting). X-ray diffraction measurements 
performed on crushed crystals reveal a pure perovskite phase free from 
secondary phases.

Pressure-dependent far-infrared reflectivity measurements at room
temperature were carried out at the infrared beamline of the
synchrotron radiation source ANKA in Karlsruhe using a Bruker IFS
66v/S Fourier transform infrared spectrometer. A diamond anvil
cell equipped with type IIA diamonds suitable for infrared
measurements was used to generate pressures up to 14.4~GPa. To
focus the infrared beam onto the small sample in the pressure
cell, a Bruker IR Scope II infrared microscope with a 15x
magnification objective was used.

The measurement of the infrared reflectivity has been performed on
the surface of as-grown BiFeO$_3$ crystals. A small piece of
sample (about 80 \nolinebreak $\mu$m $\times$ 80 \nolinebreak
$\mu$m $\times$ 40 \nolinebreak $\mu$m) was placed in the hole
(150 $\mu$m diameter) of a steel gasket. With this crystal size
and the corresponding diffraction limit, we were able to measure
reliably the frequency range above 200~cm$^{-1}$. Finely ground
CsI powder was added as a quasi-hydrostatic pressure-transmitting
medium. The ruby luminescence method \cite{Mao86} was used for the
pressure determination.

\begin{figure}[t]
\includegraphics[angle=270,width=0.95\columnwidth]{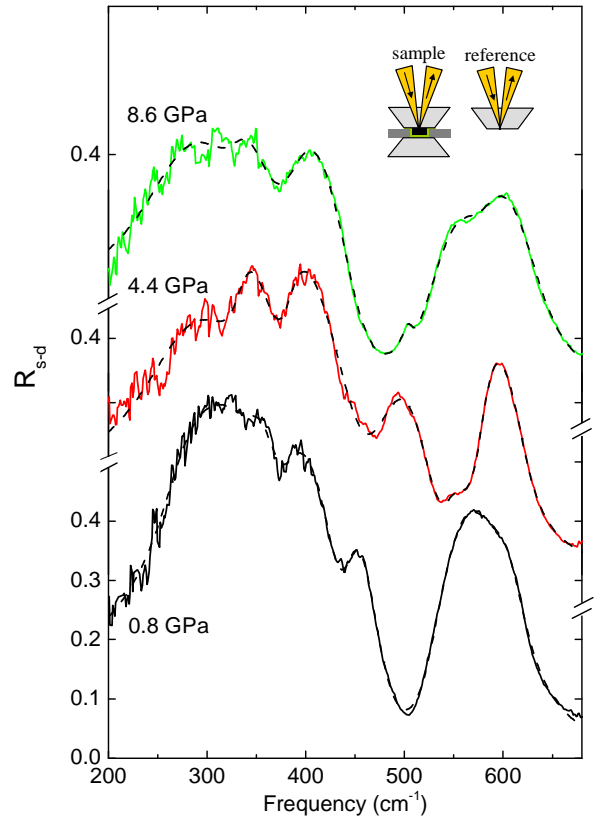}
\caption{(Color online) Room-temperature reflectivity R$_{s-d}$
spectra of BiFeO$_3$ for three selected pressures (0.8, 4.4,
8.6~GPa); the spectra are offset along the vertical axis for clarity.
The dashed lines are the fits with the generalized-oscillator
model according to Eq.~(\ref{eq:4p-fit}) (see text for details).
Inset: Measurement geometry for the reflectivity measurements, as
described in the text.} \label{fig:Rsd}
\end{figure}

Reflectivity spectra were measured at the interface between sample
and diamond. The measurement geometry is shown in the inset of
Fig.~\ref{fig:Rsd}. Spectra taken at the inner diamond-air
interface of the empty cell served as the reference for
normalization of the sample spectra. The absolute reflectivity at
the sample-diamond interface, denoted as $R_{s-d}$, was calculated
according to $R_{s-d}(\omega)=R_{\rm dia}\times
I_{s}(\omega)/I_{d}(\omega)$, where $I_s(\omega)$ denotes the
intensity spectrum reflected from the sample-diamond interface and
$I_d(\omega)$ the reference spectrum of the diamond-air interface.
$R_{\rm dia} = 0.167$ was calculated from the refractive index of
diamond, $n_{\rm dia} = 2.38$, and assumed to be independent of
pressure. This is justified because $n_{\rm dia}$ is known to
change only very little with pressure.\cite{Eremets92,Ruoff94}
Variations in synchrotron source intensity were taken into account
by applying additional normalization procedures. The
reproducibility was ensured by two experimental runs on different
crystals.

The presented high pressure spectra were collected without
polarizer, since synchrotron radiation is strongly polarized by
itself. The orientation of the samples in the pressure cell
allowed us to probe the response of the phonon modes polarized
normal to the direction of spontaneous polarization, similar to
Ref.~\onlinecite{Lobo-arxiv} (as discussed in
Section~\ref{sectionresults}).

\begin{figure}[t]
\includegraphics[angle=270,width=0.9\columnwidth]{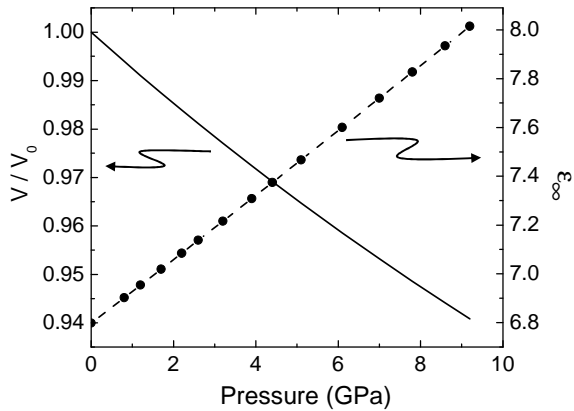}
\caption{Pressure dependence of the unit cell volume, calculated
according to the first order Birch equation of state
[Eq.~(\ref{eq:Birch})] and the high-frequency permittivity
$\epsilon_{\infty}$ as a function of pressure, calculated
according to the Clausius-Mossotti relation
[Eq.~(\ref{eq:Clausius})].} \label{fig:volume}
\end{figure}

\section{Results}\label{sectionresults}

In Fig.~\ref{fig:Rsd} we show the far-infrared reflectivity
spectra of BiFeO$_3$ at room-temperature for three selected
pressures; the spectra are offset along the vertical axis for
clarity.
Following the analysis of the infrared and tera\-hertz spectra in
Ref.~\onlinecite{Kamba07a}, we applied to our spectra the
generalized-oscillator model with the factorized form of the
complex dielectric function:\cite{Gervais83}
\begin{equation}
  \epsilon(\omega)=\epsilon_{\infty}\prod_{j=1}^{n}\frac{\omega_{LO_j}^2-\omega^2+i\omega \gamma_{LO_j}}{\omega_{TO_j}^2-\omega^2+i\omega
  \gamma_{TO_j}},
  \label{eq:4p-fit}
\end{equation}
where $\omega_{TO_j}$ and $\omega_{LO_j}$ denote the transverse
and longitudinal frequencies of the $j$th polar phonon mode,
respectively, and $\gamma_{TO_j}$ and $\gamma_{LO_j}$ denote their
corresponding damping constants. The oscillator strength
$\Delta$$\epsilon_j$ [i.e., contribution of the phonon mode to the
static dielectric constant $\epsilon(0)$] of the $j$th polar
phonon can be calculated from the formula\cite{Gervais83}
\begin{equation}
\Delta\epsilon_j =
\frac{\epsilon_{\infty}}{\omega_{TO_j}^2}\frac{\prod_{k}(\omega_{LO_k}^2-\omega_{TO_j}^2)}{\prod_{k
\neq j} (\omega_{TO_k}^2-\omega_{TO_j}^2)}. \label{eq:deps}
\end{equation}
The four-parameter oscillator model [Eq.~(\ref{eq:4p-fit})]
follows from the general properties of the dielectric function in
a polarizable lattice (pole at transverse and zero at longitudinal
eigenfrequencies of polar phonons) and it is able to describe the
permittivity of dielectrics in most cases. However, it has a
drawback since a certain combination of parameter values in
Eq.~(\ref{eq:4p-fit}) may result in unphysical values of the
complex permittivity\cite{Gervais83,Orera98} (for example,
negative losses or finite conductivity at infinite frequency).
Therefore, in our fitting procedure of the infrared reflectivity
we restricted the parameter values to those which result in an optical
conductivity vanishing at frequencies much higher than the phonon
eigenfrequencies.

\begin{figure}[t]
\includegraphics[width=0.95\columnwidth]{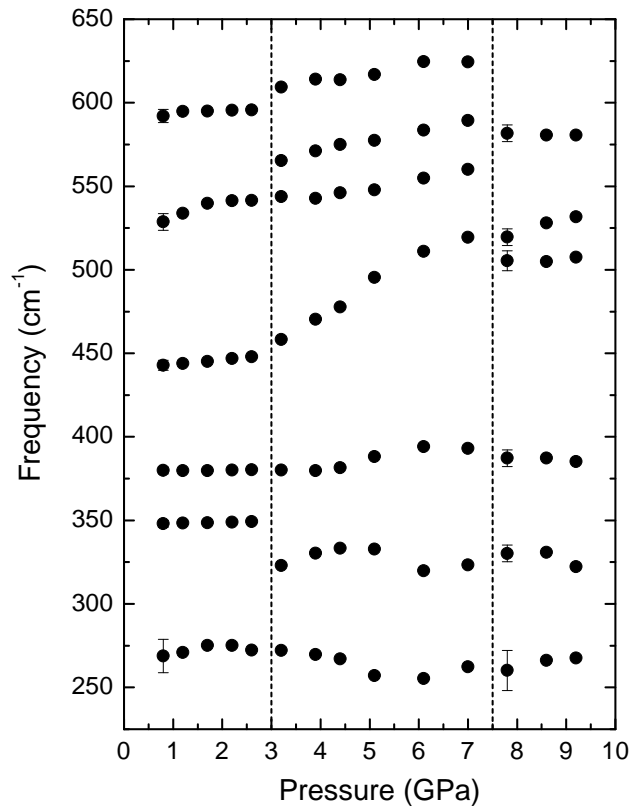}
\caption{Frequencies of the transverse optical phonons in
BiFeO$_3$ as a function of pressure, obtained by fitting the
reflectivity spectra R$_{s-d}$($\omega$) with the
generalized-oscillator model. The vertical dashed lines
indicate the pressures of the two phase transitions.}
\label{fig:omega}
\end{figure}

The dielectric function $\epsilon(\omega)$ [Eq.~(\ref{eq:4p-fit})]
is directly related to the measured reflectivity $R_{s-d}(\omega)$
at the sample-diamond interface by the Fresnel equation
\begin{equation}
R_{s-d}(\omega) =\left| \frac{\sqrt{\epsilon(\omega)}- n_{\rm dia} } {\sqrt{\epsilon(\omega)}+n_{\rm dia}}\right|^2.
\end{equation}

The pressure dependence of the high-frequency permittivity $\epsilon_{\infty}$ used in our
fitting was calculated according to the Clausius-Mossotti relation:\cite{Ashcroft76}
\begin{equation}
  \frac{\epsilon_\infty(P) - 1}{\epsilon_\infty(P) + 2} = \frac{\alpha}{3 \epsilon_0 V(P)},
  \label{eq:Clausius}
\end{equation}
where $\alpha$ is the electronic polarizability of the unit cell, which was obtained from the
lowest-pressure data. The pressure dependence of the unit cell volume, $V(P)$, has not been
measured yet; therefore we calculated $V(P)$ according to the first-order Birch equation of
state:\cite{Holzapfel96}
\begin{eqnarray}
  P(x) = \frac{3}{2}B_0 x^{-7} (1-x^2),  \label{eq:Birch}
\end{eqnarray}
where $x=[V(P)/V(0)]^{1/3}$. For the bulk modulus at zero pressure
we assumed $B_0$=130.9~GPa according to the ab-initio
calculations.\cite{Ravindran06} The resulting pressure dependence
of the unit cell volume is presented in Fig.~\ref{fig:volume}
together with the high-frequency permittivity $\epsilon_{\infty}$
as a function of pressure, calculated with
Eq.~(\ref{eq:Clausius}). The estimated value of
$\epsilon_{\infty}$ at ambient pressure is 6.8. It is higher than
the value of 4.0 reported for BiFeO$_3$ ceramics,\cite{Kamba07a}
however, lower than $\epsilon_{\infty}=9.0$ reported for single
crystals.\cite{Lobo-arxiv} Therefore, the $\epsilon_{\infty}$
value used in this work is reasonable. However, its precision is
critically dependent on several parameters which can hardly be
controlled in pressure experiments (like surface quality, parasitic
reflections from diamond anvil interfaces etc.).

The reflectivity spectra could be well fitted with the
generalized-oscillator model according to Eq.~(\ref{eq:4p-fit}).
As examples, we show in Fig.~\ref{fig:Rsd} the reflectivity
spectra R$_{s-d}$ of BiFeO$_3$ at three selected pressures and the
corresponding fits with the generalized-oscillator model. Below
$P_{c1}=3$~GPa the reflectivity spectra in the measured frequency
range can be well fitted using 6 oscillator terms. Above 3~GPa an
additional oscillator term is needed for a reasonable fit of the
spectra. Finally, above 7.5~GPa the number of oscillators reduces
to six again. The pressure dependence of the transverse phonon
frequencies is shown in Fig.~\ref{fig:omega}.

\begin{table}
\caption{\label{tab:phonons} Room-temperature fitting parameters
from Eq.~(\ref{eq:4p-fit}) to describe the reflectivity spectrum of
BiFeO$_3$ at 0.8~GPa, compared to the room-temperature parameters
obtained at ambient pressure by Lobo et al.\cite{Lobo-arxiv},
denoted by $\omega_{TO}^{amb}, \gamma_{TO}^{amb}$ and $\Delta
\epsilon^{amb}$.}
\begin{ruledtabular}
\begin{tabular}{ccccc}
$\omega_{TO} (\gamma_{TO})$ &
$\omega_{TO}^{amb}(\gamma_{TO}^{amb})$ &
$\omega_{LO}(\gamma_{LO})$ & $\Delta \epsilon$ &
$\Delta \epsilon^{amb}$ \\
\hline
269 (51) & 262 (9.1) & 348 (41) & 18.2 & 14.8\\
    & 274 (33.5) & &  & 2.45 \\
348 (36) & 340 (17.4) & 374 (43) &  0.023 & 0.27\\
380 (41) & 375 (21.6) & 433 (43) &  0.32 & 0.475 \\
443 (33) & 433 (33.8) & 472 (44) &  0.15 & 0.301\\
529 (48) & 521 (41.3) & 588 (48) &  0.69 & 1.14\\
592 (46) &   &  614 (37) &  0.019 \\

\end{tabular}
\end{ruledtabular}
\end{table}

\begin{figure}[t]
\includegraphics[angle=270,width=0.95\columnwidth]{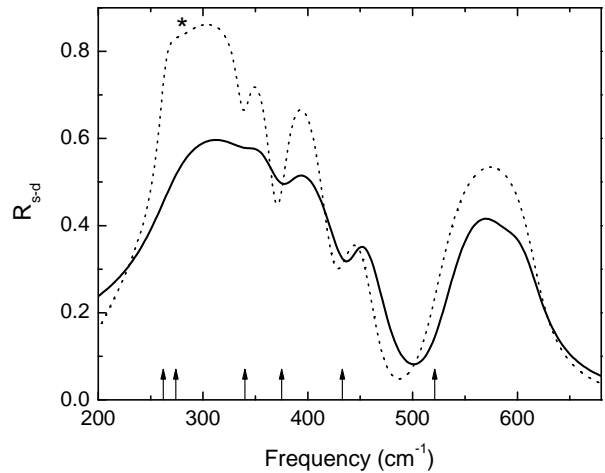}
\caption{Fit of the measured reflectivity spectrum of BiFeO$_3$ at
0.8 GPa (solid line) compared to the simulated ambient pressure
spectrum in the diamond anvil cell using the fitting parameters
from Ref.~\onlinecite{Lobo-arxiv} (dashed line). The arrows
indicate the frequencies of TO phonons found by Lobo et
al.\cite{Lobo-arxiv} The asterisk marks the kink produced by the
mode at 274~cm$^{-1}$.} \label{fig:lobo}
\end{figure}

The factor-group analysis predicts 13 infrared- and Raman-active
phonon modes for the room temperature $R3c$ phase of
BiFeO$_3$. They can be classified according to the irreducible
representations $4A_1+9E$, i.e., there are 4 $A_1$ modes polarized
along the direction of the spontaneous polarization and 9 $E$
doublets polarized normal to this direction. In addition, there are 5 $A_2$
silent modes. The frequencies of the optical phonons have been
calculated theoretically\cite{Hermet07} and determined
experimentally by infrared\cite{Lobo-arxiv} and
Raman\cite{Haumont06a,Fukumura07} spectroscopy on single BiFeO$_3$
crystals. According to the fit of our data with the
generalized-oscillator model the transverse optical modes are
located at 269, 348, 380, 443, 529 and 592~cm$^{-1}$ for the
lowest measured pressure (0.8~GPa). In Table~\ref{tab:phonons} we
list the frequencies of the transverse and longitudinal optical
modes obtained by our infrared reflectivity measurements on single
crystals at the lowest pressure together with the ambient-pressure
results for BiFeO$_3$ single crystal obtained by Lobo et
al.\cite{Lobo-arxiv} There is a very good agreement between the
transverse phonon frequencies $\omega_{TO}$ and
$\omega_{TO}^{amb}$. However, the damping constants $\gamma_{TO}$
are higher in the case of our pressure measurements. The
difference in the far-infrared reflectivity spectra R$_{s-d}$($\omega$) 
for the two sets of parameters given in Table~\ref{tab:phonons}
is illustrated in Fig.~\ref{fig:lobo}. Obviously, both
reflectivity spectra look similar and differ only in the overall
reflectivity level and the sharpness of the phonon dips. The
broadening of the phonon modes under high pressure is rather
common: it is related to the increase of the number of lattice
defects in the sample under pressure application, resulting in an
increase of the phonon scattering rate. Perhaps the mode at
274~cm$^{-1}$ which produces a small dip in the reflectivity
curve (marked by an asterisk in Fig.~\ref{fig:lobo}) observed by
Lobo et al. becomes even weaker due to the broadening effect in
our pressure measurements. Thus, it could not be reliably resolved
in the measured spectra and was therefore neglected in our fitting
procedure.

All the phonon modes listed in Table~\ref{tab:phonons}, besides
the weak mode at 592~cm$^{-1}$, belong to the $E$ representation,
i.e., they are polarized perpendicular to the direction of
spontaneous polarization [111]$_{pc}$. This indicates that the
electric field of the synchrotron radiation used in our experiment
was polarized approximately along the [-110]$_{pc}$ direction,
similar to the experiment of Lobo et al.\cite{Lobo-arxiv}

The evolution of the optical conductivity
$\sigma'(\omega)=\omega\epsilon_0\epsilon''(\omega)$ with increase
of pressure is shown in Fig.~\ref{fig:sigma1}. One can clearly see
the drastic changes of the optical conductivity spectra across the
transition pressures $P_{c1}=3$~GPa and $P_{c2}=7.5$~GPa.

\section{Discussion} \label{sectiondiscussion}

The five detected phonon modes can be assigned to the bending and
stretching modes of the FeO$_6$ octahedra, which exhibit a
displacement of the Fe$^{3+}$ cations from their centrosymmetric
position along the pseudo-cubic [111]$_{pc}$
direction.\cite{Fischer80,Kubel90} The change in the pressure
dependence of the phonon mode frequencies at $P_{c1}$ and $P_{c2}$
could thus be assigned to changes in the octahedral distortion. By
comparison with the phonon spectra of typical perovskite materials
like LaTiO$_{3}$ and BaTiO$_{3}$,\cite{Lunkenheimer03,Last57} the
experimentally observed modes can be attributed to FeO$_6$
octahedral bending and stretching modes (in the frequency ranges
200-400~cm$^{-1}$ and 400-850~cm$^{-1}$, respectively). The Bi
ions are involved only in the lower-frequency ($<$200~cm$^{-1}$)
modes located below the measured frequency range of this study.

Recent pressure-dependent Raman and x-ray diffraction studies
revealed two pressure-induced structural phase transitions at
around 3 and 10~GPa:\cite{Kreisel07a,Haumont06} The phase
transition at 3~GPa was interpreted in terms of a change of the
cation displacement and the octahedral tilting. At 10~GPa a
suppression of the cation displacements (with a concomitant
suppression of the ferroelectricity) and a change of the crystal
symmetry to orthorhombic $Pnma$, was suggested to occur. The
second phase transition is in agreement with ab-initio
calculations\cite{Ravindran06} which predict a pressure-induced
change of the crystal symmetry to the $Pnma$ group at around 13~GPa.

\begin{figure}[t]
\includegraphics[width=0.95\columnwidth]{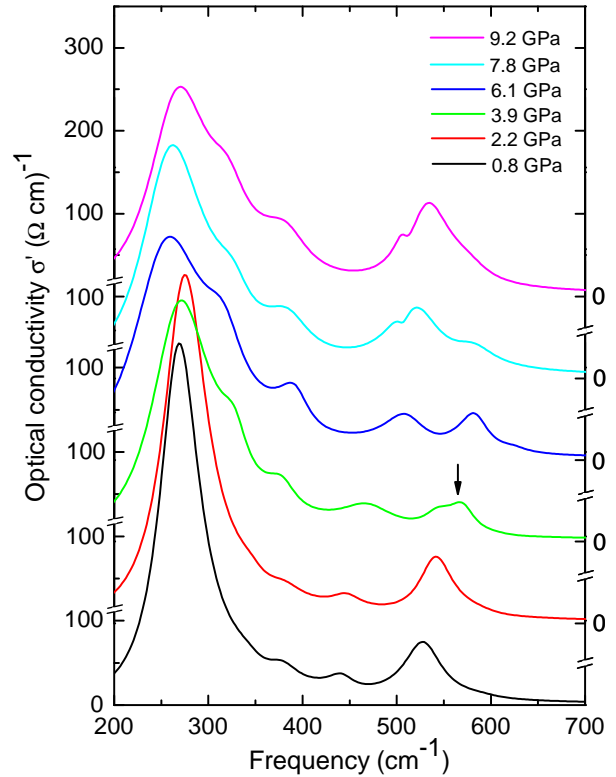}
\caption{(Color online) Real part $\sigma'(\omega)$ of the optical
conductivity of BiFeO$_3$ for selected pressures, obtained by
fitting the reflectivity spectra R$_{s-d}$($\omega$) with the
generalized-oscillator model; the spectra are offset along the vertical
axis for clarity. The arrow marks the position of the phonon
mode at 565~cm$^{-1}$ emerging above 3~GPa.} \label{fig:sigma1}
\end{figure}

Our pressure-dependent far-infrared data confirm the occurrence of
two phase transitions in BiFeO$_3$. In particular, we can confirm
the phase transition at $P_{c1}\approx$3~GPa, which is
surprisingly low considering the robustness of the
ferroelectricity in BiFeO$_3$ with respect to the temperature
increase. The high spontaneous electrical polarization 
observed also for strained BiFeO$_3$ thin
films,\cite{Wang03,Yun04,Li04} where the stress induced by the
mismatch between film and substrate is comparable to the
compressive stress produced by external pressure\cite{Haumont06}
around $P_{c1}$, suggests that the symmetry of the crystal
above 3~GPa can be still described by a polar space group, although
different from the ambient-pressure $R3c$ group. Among possible
candidates are the tetragonal $P4mm$ and the monoclinic $Cm$
groups suggested for epitaxial thin films,\cite{Singh05,Xu05}
which are energetically close to the $R3c$ structure according to
ab-initio calculations.\cite{Ravindran06} The most remarkable
signature of the phase transition at 3~GPa is the appearance of a 
phonon mode at 565 cm$^{-1}$ (see Figs.~\ref{fig:omega} and
\ref{fig:sigma1}). Furthermore, the pressure dependence of the 
frequency of the other TO phonon modes demonstrates anomalies across
the transition pressure (change of the slope of the frequency shift).
In contrast to this finding, the Raman measurements under pressure
detected the appearance of new modes and clear anomalies around 3
~GPa only for the modes below 250~cm$^{-1}$ which were not
accessible by our far-infrared study. We speculate that the new
Raman and infrared modes originate from the silent $A_2$ modes of
the parental $R3c$ phase which become active due to symmetry
change above 3~GPa. Since the activated phonon modes appear in the
frequencies ranges typical for oxygen octahedra stretching modes
as well as for modes involving vibration of the Bi ions, we
conclude that the crystal structure change at $P_{c1}$ is characterized by
the simultaneous tilting of oxygen octahedra and the Bi cations
displacement.

At higher pressures a second transition into the paraelectric
phase with $Pnma$ symmetry has been predicted
theoretically\cite{Ravindran06} at 13~GPa, and it was observed
experimentally by Raman spectroscopy and x-ray diffraction
\cite{Kreisel07a,Haumont06} at about 10~GPa. Since the unit cell
of the orthorhombic perovskite with $Pnma$ space group contains 4
formula units\cite{Iliev98}, i.e., twice more atoms than the
rhombohedral $R3c$ unit cell of the BiFeO$_3$, the number of the
phonon modes should double in the paraelectric phase. However, due
to the exclusion rule which applies to all crystals with inversion
symmetry, the Raman and infrared-active modes belong to different
symmetry species. In analogy with the perovskite
LaMnO$_3$,\cite{Iliev98} there should be in total 25 infrared
modes ($9B_{1u}+7B_{2u}+9B_{3u}$) and 24 Raman modes
($7A_g+5B_{1g}+7B_{2g}+5B_{3g}$) in the paraelectric phase of
BiFeO$_3$. The increased number of modes in the $Pnma$ phase compared
to 13 modes in the $R3c$ phase should originate from the splitting of
the $E$ symmetry doublets and the general doubling of all modes due to
the unit cell doubling. Thus, one would expect to observe a
splitting of the phonon modes across the transition pressure,
although some modes can vanish due to the selection rules. Such
effects were reported in pressure-dependent Raman
measurements of BiFeO$_3$ crystals around
9-10~GPa.\cite{Kreisel07a} Our infrared measurements demonstrate a
similar effect: above 7.5~GPa the mode at 520 cm$^{-1}$ splits
into two modes (see Figs.~\ref{fig:omega},~\ref{fig:sigma1}). On
the other hand, two of the infrared modes above 550~cm$^{-1}$
cannot be resolved above 7.5~GPa possibly as a result of exclusion
rule in the centrosymmetric $Pnma$ phase. Thus, our infrared study
confirms the pressure-induced transition into the paraelectric
phase. However, the transition pressure $P_{c2}\simeq 7.5$~GPa is
somewhat lower than the value of 9-10~GPa reported from x-ray
diffraction and Raman studies.\cite{Kreisel07a,Haumont06} This
difference in pressure can be understood by the different
pressure transmitting media used in the two experimental investigations
(argon in the earlier measurements\cite{Kreisel07a,Haumont06} and
CsI in our case), since under more hydrostatic conditions (argon)
the transition is expected to occur at {\it higher}
pressure.\cite{Frank06}

\section{Summary}
We have studied the far-infrared reflectivity of the multiferroic
material BiFeO$_3$ under high pressure. The frequencies of the
transverse optical phonons demonstrate two distinct anomalies in
their pressure dependence at 3.0 and 7.5~GPa, which can be assigned
to structural phase transitions. The results of our infrared
spectroscopy study are in good agreement with recent Raman and
x-ray diffraction studies under
pressure.\cite{Kreisel07a,Haumont06} The analy\-sis of the phonon
behavior suggests that the transition at 3~GPa is characterized by
the simultaneous tilting of oxygen octahedra and the Bi cations
displacement. The changes across 7.5~GPa are consistent with a
transition into the paraelectric $Pnma$ phase predicted
theoretically\cite{Ravindran06} and observed experimentally by a
recent x-ray diffraction study.\cite{Kreisel07a}

\subsection*{Acknowledgements}
We acknowledge the ANKA Angstr\"omquelle Karlsruhe for the
provision of beamtime and we would like to thank B. Gasharova,
Y.-L. Mathis, D. Moss, and M. S\"upfle for assistance using the
beamline ANKA-IR. Financial support by the Bayerische
Forschungsstiftung and the DFG through the SFB 484 is gratefully
acknowledged. Support from the French National Research Agency
("ANR Blanc") is greatly acknowledged by RH and JK. JK thanks the
European network of excellence FAME and the European STREP
MaCoMuFi for financial support.

\end{document}